\documentclass[12pt,preprint]{aastex}
\newcommand{\etal}{\mbox{\rm et al.}}

\newcommand{\msun}{\mbox{$M_{\odot}$}}

\newcommand{\mearth}{\mbox{$M_\oplus$}}

\newcommand{\teff}{\mbox{$T_{\rm eff}$}}
\newcommand{\fe}{\mbox{\rm [Fe/H]}}

\newcommand{\rhk}{\mbox{$\log R^\prime_{\rm HK}$}}
\newcommand{\shk}{\mbox{$S_{\rm HK}$}}

\newcommand{\mjup}{\mbox{$M_{\rm Jup}$}}
\newcommand{\rchisq}{\mbox{$\chi_{\nu}^2$}}     % reduced chi-square
\newcommand{\msini}{\mbox{$M\sin i$}}           % planet mass
\newcommand{\ms}{\mbox{m s$^{-1}$}}

\newcommand{\kms}{\mbox{km s$^{-1}$}}
\newcommand{\rms}{\mbox{rms}}

\newcommand{\snr}{\mbox{\rm signal-to-noise ratio}}
\newcommand{\caii}{\ion{Ca}{2} H \& K}
  
\received{}
\accepted{}
  
\slugcomment{to appear in PASP}
  
\shortauthors{Apps \etal}
\shorttitle{HIP~79431}
\begin{document}
  
\title{M2K: I. A Jovian Mass Planet around the M3V star HIP~79431\altaffilmark{1}}
\author{Kevin Apps\altaffilmark{2},
Kelsey I. Clubb\altaffilmark{3},
Debra A. Fischer\altaffilmark{3,4},
Eric Gaidos\altaffilmark{5},
Andrew W. Howard\altaffilmark{6},
John A. Johnson\altaffilmark{7},
Geoffrey W. Marcy\altaffilmark{6},
Howard Isaacson\altaffilmark{6},
Matthew J. Giguere\altaffilmark{4},
Jeff A. Valenti\altaffilmark{8},
Victor Rodriguez\altaffilmark{3},
Carley Chubak\altaffilmark{6}, 
\and Sebastien Lepine\altaffilmark{9}
}
  
\email{debra.fischer@yale.edu}

\altaffiltext{1}{Based on observations obtained at the W. M. Keck Observatory,
which is operated by the University of California}
  
\altaffiltext{2}{75B Cheyne Walk, Surrey, RH6, 7LR, United Kingdom}

\altaffiltext{3}{Department of Physics \& Astronomy, 
San Francisco State University,
San Francisco, CA  94132, USA}

\altaffiltext{4}{Department of Astronomy, 
Yale University, New Haven, CT 06511, USA}

\altaffiltext{5}{Department of Geology and Geophysics, University 
of Hawaii, Honolulu, HI 96822, USA}
  
\altaffiltext{6}{Department of Astronomy, 
University of California, Berkeley, Berkeley, CA 94720, USA}

\altaffiltext{7}{Department of Astronomy, 
California Institute of Technology, Pasadena, CA 91125, USA} 

\altaffiltext{8}{Space Telescope Science Institute, 3700 San Martin Dr. 
Baltimore, MD 21218} 

\altaffiltext{9}{American Museum of Natural History
New York, NY 10023, USA}

\begin{abstract}
Doppler observations from Keck Observatory reveal the presence of 
a planet with \msini\ of 2.1 \mjup\ orbiting the M3V star HIP~79431.
This is the sixth giant planet to be 
detected in Doppler surveys of M~dwarfs 
and it is one of the most massive planets discovered around an 
M~dwarf star.  The planet has an orbital
period of 111.7 days and an orbital eccentricity of 0.29. 
The host star is metal rich, with an estimated \fe\ = +0.4. 
This is the first planet to emerge from our new survey of 
1600 M-to-K dwarf stars. 

\end{abstract}


\section{Introduction}
Doppler surveys for exoplanets are monitoring almost every bright,
chromospherically inactive solar type star (late F, G and early
K types) and a fraction of the known M~dwarfs within 30 parsecs. These
successful programs have discovered about 400 exoplanets ranging
in mass from a few times that of the Earth up to the lower mass limit
for brown dwarfs (i.e., about 12 \mjup).  This ensemble of exoplanets
has revealed correlations between both the metallicity and mass of the
host star with the occurrance rate of exoplanets \citep{sim04, fv05,
lm07, j07}.

The occurrence rate of gas giant planets around M~dwarfs appears to be lower than 
for F, G and early K type stars. \citet{e03} detected only one planetary companion 
in a sample of 100 M dwarf primaries.  \citet{j07} carried out a statistical 
analysis and showed that late K and early M~dwarf stars have a $1.8 \pm 1.0$\% 
occurrence rate for Jovian planets compared to $4.2 \pm 0.7$\% for solar-mass 
stars and $8.9 \pm 2.9$\% for higher mass subgiants. 

However, the protoplanetary disks of M~dwarfs may still be a robust environment 
for the formation of lower mass planets.  \citet{f09} find that 30\% of 
the twenty known planets with $\msini < 0.1 \mjup$ orbit M dwarf stars.
Consistent with this result, microlensing surveys
\citep{sumi09} find far more planets with masses in the 
range between a few to 20 \mearth.
These empirical results are compatible with core accretion 
models of planet formation. \citet{il05} predict that gas accretion will 
be quenched by gap formation in disks with smaller aspect ratios; the low 
surface density in the protoplanetary disks of M~dwarfs implies a low 
formation probability for gas giants because massive cores don't form 
before gas depletion. \citet{l04} also surmise that a significant 
population of ``failed Jupiters'' with cores of a few Earth masses should exist 
around M~dwarfs. 

M~dwarfs constitute the majority of stars populating our galaxy. 
Among the $\sim 150$ stars within 8 parsecs, about 120 are early-type 
M~dwarfs, while only 15 are G dwarfs. 
Low mass stars are attractive targets for searches of terrestrial mass
planets within the habitable zone where liquid water might be present 
\citep{t07, g07}. The habitable zone is closer to 
low luminosity stars and the shorter period planets induce larger, 
more easily detected reflex velocities. 
Yet, fewer than ten percent of late K and early M dwarf stars within 30 parsec 
are being monitored by Doppler surveys \citep{m09, b06, e03}. These 
stars are challenging targets because they are intrinsically faint
at optical wavelengths. Indeed, many late type M~dwarfs beyond ten parsecs 
are only now being detected; recently, the 
SUPERBLINK survey \citep{l05, ls05} identified a few thousand M~dwarfs closer
than 30 parsecs, 
opening up the possibility for larger scale surveys of these stars.

\section{A New M and K dwarf Survey} 

To learn more about the rate of exoplanet occurrence for M~dwarf stars 
and mass and stellar metallicity dependences, we have started a survey of 1600 late K and 
early M dwarfs. The target stars are drawn from the SUPERBLINK survey, which 
provides an all-sky census of stars with proper motion larger than 
40 $mas\ yr^{-1}$ down to a magnitude limit $R\approx19.5$. Main sequence 
stars are identified based on their location in a reduced proper motion 
diagram. A fraction of the stars are listed in the Hipparcos catalog; for 
those stars we calculate the distance from their Hipparcos parallax. All 
other stars have their distances estimated based on the empirical $V,V-J$ 
color-magnitude relationship from \citet{l05}. We select as probable 
late K and M dwarfs all stars with optical-to-infrared color $V-J > 2.75$, 
which roughly corresponds to absolute visual magnitudes $M_V > 8.5$ and 
masses $M < 0.6 M_{\odot}$. We further narrow down our sample to stars with 
photometric distances $d < 50$ pc and visual magnitudes $V < 12.5$. The
low proper motion limit and short distance range minimizes the kinematic 
bias, and the sample is expected to be complete for stars with transverse 
velocities $v_T>9.5\ \kms$ ($v_T>5.7\ \kms$ for the 30 pc subsample), 
which effectively samples the local disk population over the full range 
of typical stellar ages, as the selection even includes stars from nearby 
young moving groups \citet{ls09}. Of the final sample of 1600 stars, 
40\% are listed in the Hipparcos catalog.

We carry out low resolution spectroscopic screening of all candidates before 
they are observed at Keck using the Mark III spectrograph at the 1.3-m 
McGraw-Hill Telescope at teh MDM Observatory at Kitt Peak. These observations 
are used to confirm the spectral type and main sequence status before 
beginning observations at Keck. 

Modeled after the N2K program \citep{f05}, our program uses a quick-look 
strategy to obtain three or four nearly consecutive observations of target stars 
to flag short-period candidates. We then continue to obtain approximately 
log-spaced time series observations to detect longer period planets. Here, 
we present the first planet to emerge from this program. 

\section{HIP 79431}
  
\subsection{Stellar Characteristics}
  
HIP~79431 is an M3V star with apparent magnitudes of $V = 11.34$,
$J = 7.56$, $H = 6.86$, $K = 6.59$ and color \bv\ = 1.486. 
The {\it Hipparcos} parallax \citep{esa97, vanleeuwen07} 
of $69.46 \pm 3.12$ mas yields a distance of 
14.4 parsecs and we use this to calculate the absolute magnitudes
$M_V = 10.47$, $M_J = 6.69$, $M_H = 5.99$, $M_K = 5.72$. The proper 
motions $\mu_{RA} = 28.31\ {\rm mas\ yr^{-1}}$, $\mu_{Dec} = -208.36\ {\rm mas\ yr^{-1}}$ 
measured by the {\it Hipparcos} mission \citep{esa97} and our measured absolute
radial velocity of -4.7 \kms\ correspond to the small space velocities 
of $U = +8.5\ \kms$, $V = +2.1\ \kms $, $W = -4.8\ \kms$. 

\citet{d00} have established mass-luminosity calibrations for 
low mass stars and find low dispersion relations for $M_J$, 
$M_H$ and $M_K$.  The average of these three infrared mass-luminosity 
relations yield an average stellar mass of 0.49 \msun\ for HIP~79431.

It is challenging to assess metallicity from M dwarf spectra, 
largely because of uncertainties in the molecular line data.
\citet{bo05} first derived a photometric calibration for M~dwarf metallicities
and this work has been updated by \citet{ja09}. Both groups used spectral 
synthesis modeling to determine \fe\ for bright F and G type stars 
in binary systems with M dwarf companions 
and assigned the metallicity of the primary star to the 
M dwarf companion. Based on the star's height above the $M_K$ vs $V-K$ 
main sequence, the calibration of \citet{ja09} yields a metallicity of 
$\fe = +0.4 \pm 0.1$.

The effective temperature for M~dwarfs is usually derived from 
a color-temperature relation. The K-band temperature relation 
calibrated by \citet{c08} is insensitive to metallicity and M dwarf 
fluxes peak near the infrared $K$ band. The \citet{c08} relation 
has a scatter of only 19 K and yields \teff = 3191 K for 
HIP~79431. This is similar to the effective temperature of 3236 K 
that can be derived with the less robust \bv\ calibration of 
\citet{vf05} for FGK type stars.

Because the continuum flux near the \caii\ lines is so low 
in M~dwarfs it is typical to see strong emission from the core of the 
\caii\ lines. \citet{if09} measure $S_{HK} = 1.15 \pm 0.06$ in this star. 
The \rhk, activity-based rotation period and ages are not calibrated for stars 
with \bv\ greater than 1.0, however \citet{if09} find that this value of 
$S_{HK}$ is in the 50th percentile for stars of this color, so the 
star does not have an unusually high level of chromospheric activity for  
a dwarf star with this \bv\ color.
The stellar characteristics are summarized in Table 1. 

\subsection{Doppler Observations and Keplerian Fit}
HIP~79431 was added to the Keck program in April 2009 as part of our exoplanet 
survey of low mass stars. We obtained 13 Doppler measurements 
of HIP~79431 over six months using the HIRES spectrometer \citep{v94}. 
Exposure times of 600 seconds for this $V=11.3$ star 
yielded a \snr\ of just under 100. 

Our Doppler analysis makes use of an iodine absorption cell in the 
light path before the entrance slit of the HIRES spectrometer.  The iodine 
absorption lines in each program observation are used to model the  
wavelength scale and the instrumental profile of the telescope and
spectrometer optics for each observation \citep{m92, b96}. The typical 
Doppler precision for this faint star is about 3 \ms. Based on the \rms\ scatter
of other stars on our program, we have empirically derived an additional 
error which may arise from astrophysical noise sources or systematic 
uncertainties in our data. Figure 1 shows time series Doppler observations 
for stars with constant radial velocities and \bv\ colors similar to HIP~79431. 
The typical \rms\ scatter is about 2.5 \ms.  We adopt 2.5 \ms\ as representative 
of the combined radial velocity jitter and systematic errors for late type 
stars and add this in quadrature with our formal single measurement errors 
when fitting and plotting our data. 

The observation dates, radial velocity data and measurement 
uncertainties for HIP~79431 are listed in Table 2.  In this Table, 
the measurement uncertainties are not increased by the 2.5 \ms\ jitter term. 
The radial velocities exhibit an unambiguous signal. The radial velocities
were fit with a Keplerian model and the uncertainties in the orbital parameters
were derived using a bootstrap Monte Carlo analysis with 1000 realizations 
of the data \citep{ma05} in the following way: first, the theoretical velocities 
from the best-fit Keplerian model are subtracted, then the residual velocities 
are randomized, added back to the theoretical velocities and refit. This 
analysis captures the full range of uncertainties, including systematic errors, 
however, it will also scramble a signal from any additional planets, so it is 
only appropriate for good \rchisq\ fit models. 

The Keplerian fit yields an orbital period of $111.7 \pm 0.7$ days and a 
semi-velocity amplitude of $149.5\ \pm 2.5\ \ms$. The stellar mass of 0.49 \msun\ 
implies a companion mass with \msini\ = 2.1 \mjup\ and a semi-major axis of 0.35 AU. 
The parameters for the Keplerian orbit are listed in Table 3. 

The Doppler velocities are plotted in Figure 2 with a solid line indicating 
the best fit Keplerian model. In this plot, the single measurement errors have 
been increased by adding a modest 2.5 \ms\ for stellar jitter and systematic 
errors. The \rms\ of the residuals to our Keplerian model is 
3.9 \ms\ and the \rchisq\ fit is 0.84, demonstrating that the fit is consistent 
with our single measurement errors and the added assumed jitter of 2.5 \ms.
As such a good fit implies, there is no evidence for additional planets. 

\section{Planet or Brown Dwarf?}

Because the Doppler technique derives \msini\ and not the total mass,
this leaves open the question of whether this is actually a stellar
binary system in a nearly face-on orbit.  HIP~79431 is only 14.4~pc
away and a stellar companion on a nearly face-on orbit would induce
motion around the barycenter of order 10~mas.  The re-analysis of the
{\it Hipparcos} astrometry is consistent (goodness-of-fit parameter of
0.15) with the proper motion of a single star \citep{vanleeuwen07}.
We used the intermediate astrometric data \citep{esa97} to place an
upper limit on the companion mass as a function of confidence level
(one minus the false alarm probability FAP). The abscissa residuals to
the {\it Hipparcos} parallax and proper motion solution span about 10 orbits
and are plotted in Figure 3.  There is no power in a Lomb-Scargle periodogram 
near the orbital period of 111 days in these data.  Assuming gaussian
statistics, a companion more massive than 0.3 $M_{\odot}$ can be ruled
out with 90\% confidence (Figure 4), but, the {\it Hipparcos} data by
themselves cannot eliminate the possibility of a late-type M or brown
dwarf companion.  We can, however, rule out the possibility that the
location of HIP~79431 above the main sequence is due to a nearly
equally-luminous companion and therefore conclude that this location is consistent 
with the high stellar metallicity \citep{ja09}.

Even a brown dwarf mass would require an improbable nearly 
face-on orbit.  If we take the lower limit for stellar mass as 70 \mjup\ 
then the maximum inclination $i$ for our system to surpass the stellar mass threshold
is given by:

\begin{equation}
i_{70\ Jup} = \arcsin{\msini \over 70\ \mjup},
\end{equation}

The probability that this particular inclination was observed can 
be calculated under the assumption that the orientation of orbits 
is random as: 

\begin{equation}
P(star) = \cos(0) - \cos(i_{70\ Jup}) = 4 \times 10^{-4},
\end{equation}

\noindent
where $i$ is the orbital inclination from Eqn (1).
We can likewise calculate the probability that HIP~79431b falls 
in the brown dwarf range.  This time, the inclination limit is far  
less severe since the planet only needs to have a true mass 
of 12 \mjup: 

\begin{equation}
i_{12\ Jup} = \arcsin{\msini \over 12 \mjup}
\end{equation} 

\begin{equation}
P(brown\ dwarf) = \cos(i_{70\ Jup}) - \cos(i_{12\ Jup}) = 0.015
\end{equation}

So, in order for HIP~79431b to be a stellar companion, an inclination
of less than $2^\circ$ from face-on is required and that orientation
has a probability of only 0.04\%.  If HIP~79431b is really a brown
dwarf companion, the orbital inclination must be between $2^\circ$ - 
$10^\circ$; the probability of this orientation is 1.5\%. 

\section{Transit Ephemeris}

Because HIP~79431b has a semi-major axis of 0.36 AU, the planet has a 
relatively low transit probability. The planet also resides in 
a moderately eccentric orbit that brings it as close as 0.25 AU to the 
host star at periastron. For the orbit of HIP~79431b, 
$\omega = 288 \pm 3$ so a primary transit would occur when the planet 
is close to apastron.  This further reduces the probability 
of a primary transit and we calculate a transit probability of only 0.5\%. 
In this system, the secondary eclipse is somewhat more likely; without 
prior knowledge of a primary transit, we calculate a probability of 
0.85\% for the secondary eclipse.  

If a transit were to occur, the transit depth would be remarkable.
A planet with a mass of 2.1 \mjup\ is supported by electron 
degeneracy and has a radius that is only a factor of four times smaller 
than the M dwarf host. We calculate a prospective primary transit 
depth of 98 millimags and transit 
duration of almost 7 hours. The probability for a secondary eclipse 
is about double that for the primary transit, however the photometric 
decrement would only be about 0.004 millimags.

Prospective transit times were predicted from the best-fit Keplerian parameters. 
The Thiele-Innes constants were used to project 
the orbit onto the plane of the sky and to determine the true anomaly 
corresponding to the central transit time, $T_c$. The time of the next
$T_c$ is calculated to be 2010 Feb 21 at UT 18:17:23:92 with an uncertainty 
of 1.6 hours. The uncertainty in $T_c$ was determined by fitting the 1000 
trials of Keplerian fits (used to derive uncertainties in the orbital parameters). 
For each trial, the central transit time, ingress and egress were 
calculated and the standard deviation provided an estimate of the uncertainty. 
Future transit times can be estimated by rolling forward an 
integral number of orbital periods, however uncertainty in $T_c$ increases 
with time from our last Doppler measurement. 
  
\section{Discussion}
As part of a new Doppler quick-look survey for late K and early M~dwarfs, 
we have detected a planet with \msini\ = 2.1 \mjup\ orbiting the 
M3V star, HIP~79431.  The host star appears to be metal-rich with a 
photometrically-determined \fe = +0.4. 

Massive planets around late type stars are easy detections. 
However, the sample of M~dwarf stars with planets is relatively small; 
only six have detected gas giant planets ($\msini > 0.5 \mjup$).
The remarkable system of planets orbiting GJ~876 (M4V) has three planets and two
of these are gas giants in a 2:1 mean motion resonance; GJ~876b 
has \msini\ = 1.935 \mjup\ and orbital period of about 30 d 
and GJ~876c has \msini\ = 0.619 \mjup\ with an orbital period 
of 60 d \citep{m01, m98, d98}. GJ~849 (M3.5V) has a planet 
with \msini = 0.82 \mjup\ in a 5-year orbit with significant 
residual velocities that now exhibit curvature, consistent with an additional planet 
in a wider orbit. 
GJ~317 (M3.5V) has a planet 
with \msini\ = 1.17 \mjup\ with a period of about 1.9 years and residual velocities 
consistent with Jupiter-mass planet in an orbit of more than seven years \citep{j07}. 
GJ~832 (M1.5V) has a Jupiter-like planet in a 9.4-year orbit \citep{b09}. 
GJ~179 (M3.5V) has a Jupiter-mass planet in a 6.3 year orbit \citep{how09}.
GJ~649 (M1.5V) has a Saturn-mass planet in a 1.6 year orbit \citep{j09}.

It is curious that half of the known giant planets orbiting low 
mass M~dwarf stars reside in relatively wide orbits. \citet{kk09} find 
that grain growth time scales are limited by photo evaporation with 
more rapid loss of disks in high mass stars. They note that the stellar mass-dependent
disk dispersal timescales may account for the fact that planets orbiting higher-mass 
stars have wider orbits, since the disk may disperse more rapidly and halt 
inward migration. Perhaps lower surface density creates an analogous migration-hindering 
environment for low mass stars. Or, perhaps the protoplanetary disk 
of M stars forms fewer planets so that planet-planet scattering events are less 
common. With less competition for raw materials in the disk, it might be possible 
for single or well-spaced giant planets to dominate at wide separations.  
This is also a domain where gravitational instability could play a role {boss06}.

HIP~79431b is one of the most massive planets detected in orbit around an M dwarf 
star.  The calculated probability that this is actually 
an equal mass stellar companion is vanishingly small: $4 \times 10^{-4}$. The 
The {\it Hipparcos} astrometric data do not rule out the 
possibility that this is a low mass star or brown dwarf companion, but 
we calculate only a 1.5\% probability that HIP~79431b has an orbital 
inclination less than $10^\circ$ and a true mass greater than 12 \mjup. 
Statistically, this leaves a 98.5\% probability that HIP~79431b is a 
planet with a mass between 2.1 and 12 \mjup. 

The Keplerian motion induced by a single planet can explain our data,
although much less massive planets may exist on interior or exterior
orbits. The excellent agreement of the single planet Keplerian planet solution
with the data limit the mass of any second planet in the system.  We
performed a linear perturbation analysis around the one-planet
solution and found that we can rule out planets more
massive than $M_3 \sin i$ of 12 or 15 Earths near the 3:1 or 2:1
interior resonances, respectively.

Although Doppler surveys show that massive companions to M~dwarf 
stars are rare, evidence for a few Jovian-mass planets candidates has emerged 
from microlensing surveys that sample wider orbits. 
(Because low mass stars dominate the galactic population, they are
common lenses for background stars.) 
\citet{g02} place an upper limit and find that fewer than
33\% of M~dwarfs have Jupiter-mass companions between 1.5 and 4 AU.
Nevertheless, only a small number
of Jovian-mass planets have been detected in microlensing surveys
\citep{b04, u05, g08, d09}.

\acknowledgements
We gratefully acknowledge the dedication and support of the Keck
Observatory staff, especially Grant Hill and Scott Dahm for support 
of HIRES and Greg Wirth for support of remote observing. 
DAF acknowledges research support from NASA grant
NNX08AF42G as well as NASA support through the Keck PI data analysis 
Fund. EG acknowledges support by NSF grant AST0908419.
AWH gratefully acknowledges support from a Townes
Postdoctoral Fellowship at the U.C. Berkeley Space Sciences
Laboratory.  

Data presented herein were obtained at the W. M. Keck 
Observatory from telescope time allocated to the National Aeronautics
and Space Administration through the agency's scientific 
partnership with the California Institute of Technology and the 
University of California.  We also thank the TAC of the University 
of Hawaii for providing time for this project. 
The Observatory was made possible by 
the generous financial support of the W. M. Keck Foundation.
The authors wish to recognize and acknowledge the very significant 
cultural role and reverence that the summit of Mauna Kea has always 
had within the indigenous Hawaiian community. We are most fortunate
to have the opportunity to conduct observations from this mountain. 
This research has made use of the SIMBAD database, operated
at CDS, Strasbourg, France, and of NASA's Astrophysics Data System
Bibliographic Services.

\clearpage

\clearpage  

\begin{figure}
\plotone{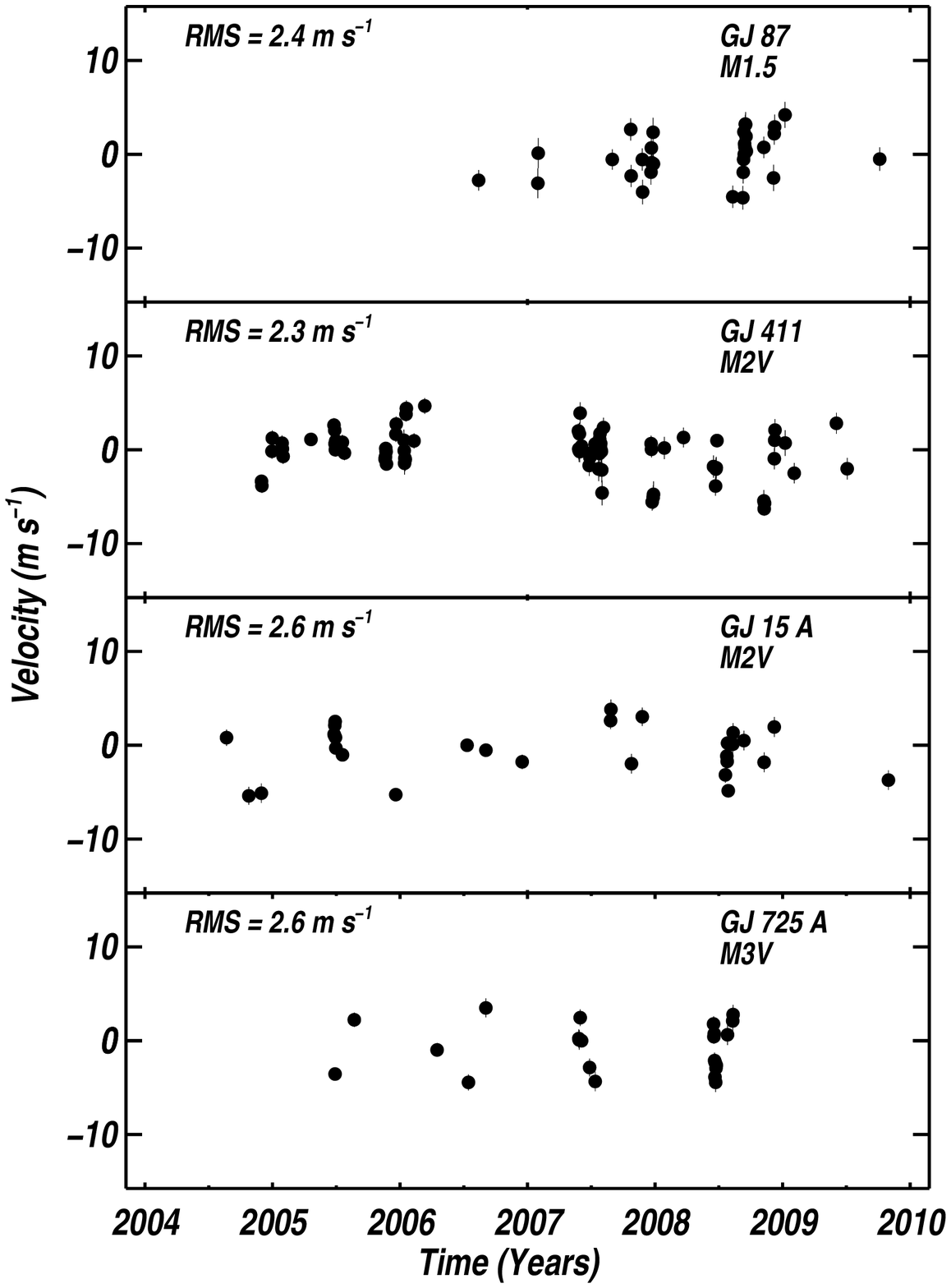}
\figcaption{Doppler measurements of four M~dwarf stars with 
constant radial velocities. These stars have an \rms\ scatter of 
about 2.5 \ms\ and represent velocity precision typical of 
late type stars on our program. }
\end{figure}

\begin{figure}
\plotone{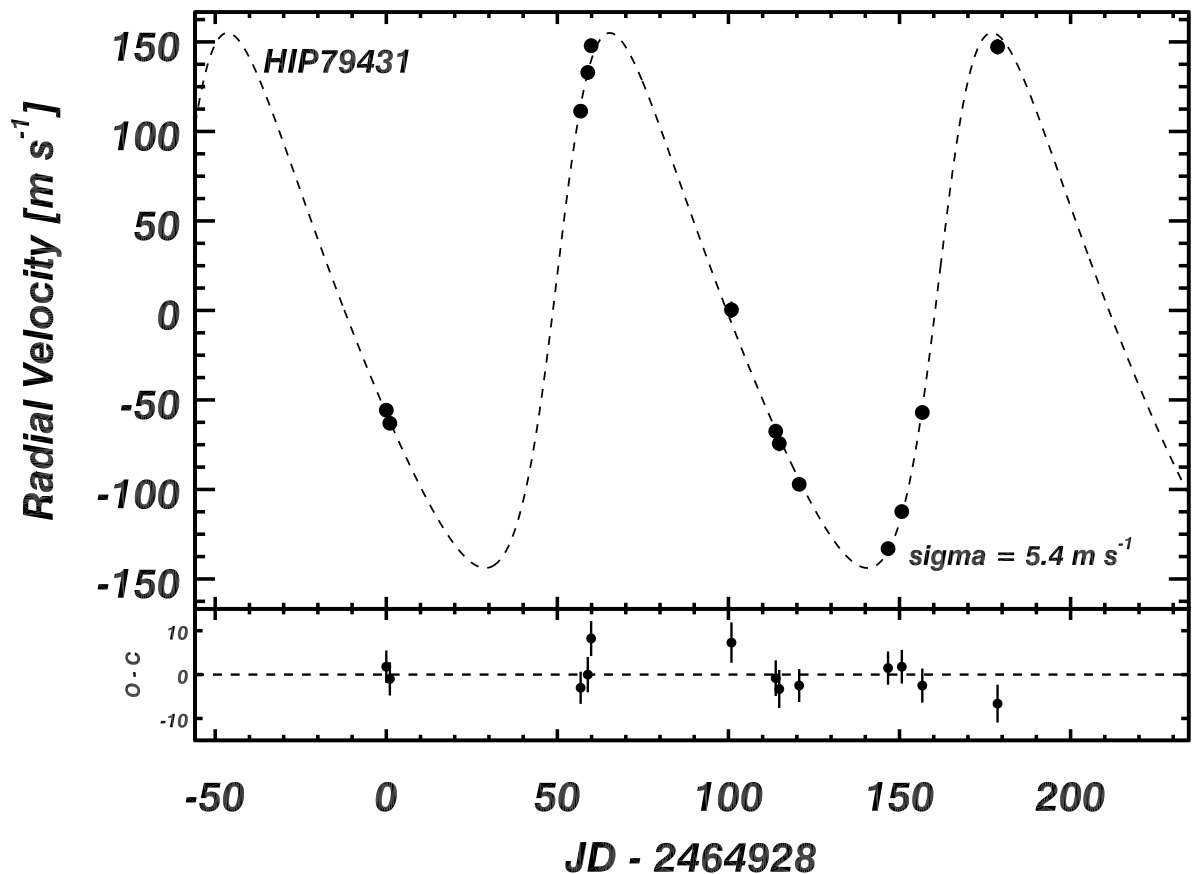}
\figcaption{Time series radial velocities from Keck 
Observatory are plotted for HIP~79431 with
2.5 \ms\ of expected velocity jitter added in 
quadrature with the single measurement uncertainties. 
The Keplerian model is overplotted with 
an orbital period of 111 days, velocity amplitude 
of 149.5 \ms\ and eccentricity, $e = 0.29$. 
With these parameters and the stellar mass of 0.49 \msun, 
we derive a planet mass, \msini\ = 2.1 \mjup\ and semi-major
axis of 0.36 AU. }
\end{figure}

\begin{figure} 
\plotone{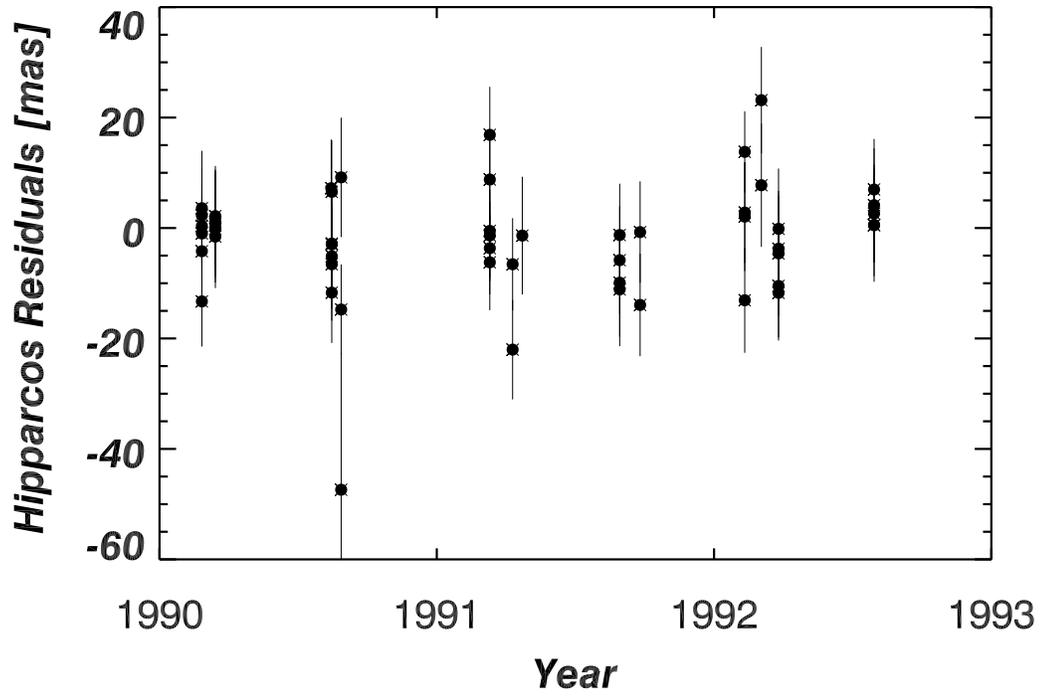}
\figcaption{Astrometric residuals (unbinned) from the {\it Hipparcos} intermediate 
data after fitting for parallax and proper motion.}
\end{figure}

\begin{figure}
\plotone{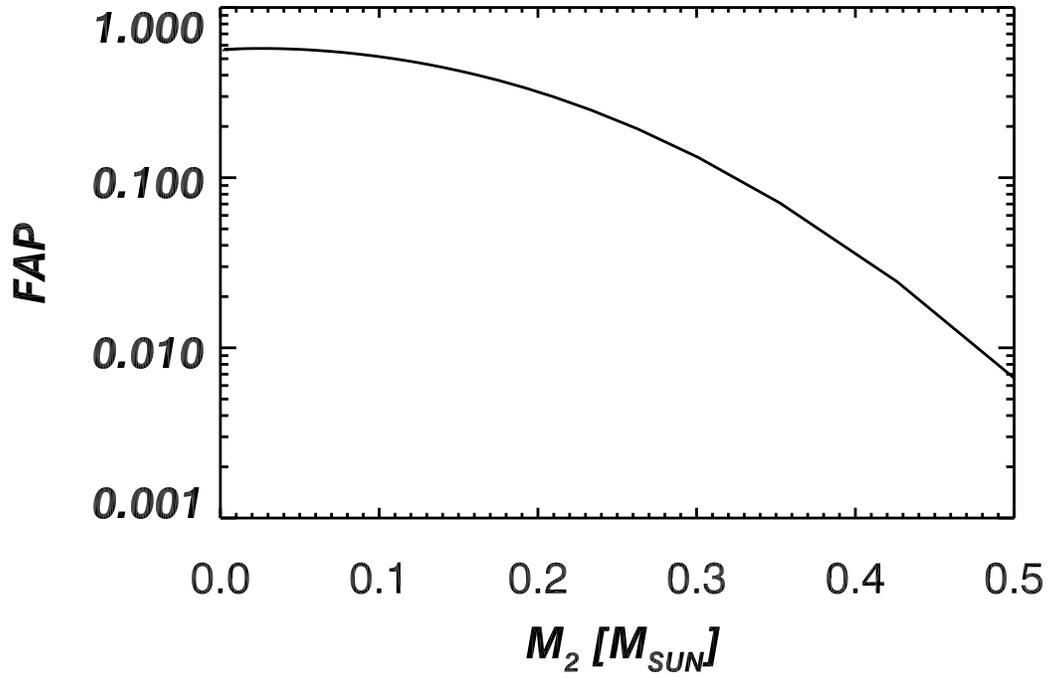}
\figcaption{False alarm probability (one minus the confidence level) of the companion exceeding a given mass given the {\it Hipparcos} abscissa residuals for HIP~79431}
\end{figure}

%\begin{figure} 
%\plotone{figs/f4.eps}
%\figcaption{If the orientation of an eccentric orbit has periastron passage along the 
%line of sight, the transit probability increases.  However, in the orientation 
%of the orbit for HIP~79431b apastron passage is oriented toward the Earth 
%resulting in a low transit probability.}
%\end{figure}

\clearpage

\begin{deluxetable}{ll}
\tablenum{1}
\tablecaption{Stellar Parameters: HIP~79431}
\tablewidth{0pt}
\tablehead{\colhead{Parameter}  & \colhead{}  \\
}
\startdata
$V$          &   11.34    \\
$M_V$        &   10.47    \\
$M_K$        &    5.72    \\
\bv          &    1.486    \\
Spectral Type    &    M3V     \\
Distance [pc] &  14.4     \\
\teff        &  3191 (1000 \\
\fe          &   +0.4 (0.1)  \\
$M_{star}$    &   0.49 (0.05) \\
\shk         &  1.15 (0.06)  \\
Radial Velocity [\kms]  & -4.7 \\
\enddata                        
\end{deluxetable} 

\begin{deluxetable}{rrcc}
\tablenum{2}
\tablecaption{Radial Velocities for HIP~79431}
\tablewidth{0pt}
\tablehead{ 
    \colhead{}   &
    \colhead{RV} & 
    \colhead{$\sigma_{\rm RV}$} \\
       \colhead{JD-2440000} & 
       \colhead{(\ms)} & 
       \colhead{(\ms)} \\ 
 }
\startdata
    14928.07496  &    -63.11  &      2.73 \\ 
    14929.11611  &    -70.45  &      2.84 \\ 
    14984.87477  &    103.92  &      2.68 \\ 
    14986.96264  &    125.53  &      3.14 \\ 
    14987.94866  &    140.47  &      3.10 \\ 
    15028.96005  &     -7.06  &      3.88 \\ 
    15041.89538  &    -74.88  &      3.21 \\ 
    15042.92741  &    -81.75  &      3.55 \\ 
    15048.75289  &   -104.55  &      2.80 \\ 
    15074.72733  &   -140.47  &      2.84 \\ 
    15078.74506  &   -119.78  &      2.90 \\ 
    15084.73424  &    -64.40  &      2.97 \\ 
    15106.73408  &    139.86  &      3.52 \\ 
\enddata
\end{deluxetable}
\clearpage

\begin{deluxetable}{ll}
\tablenum{3}
\tablecaption{Orbital Parameters for HIP~79431b}
\tablewidth{0pt}
\tablehead{\colhead{Parameter}  & \colhead{} \\
} 
\startdata
Period (days)            &  111.7 (0.7)     \\
${\rm T}_{\rm p}$ (JD)   &  2454980.3 (1.2) \\
$\omega$ (deg)           &  287.4 (3.2)     \\
Eccentricity             &  0.29 (0.02)     \\
K$_1$ (\ms)              &  149.5 (2.5)     \\
$a_{rel}$ (AU)           &  0.36            \\
$M\sin i$ (M$_{Jup}$)    &  2.1             \\
${\rm Nobs}$             &  13              \\
RMS (\ms)                &  3.9             \\
\rchisq\                 &  0.84            \\
\enddata                        
\end{deluxetable}                          
\clearpage

\end{document}